\documentclass[prd,preprint,nofootinbib,nobibnotes]{revtex4}
\usepackage{epsf,epsfig}
\usepackage{amssymb,amsmath,amsfonts}

\newcommand{\eqn}[1]{(\ref{#1})}
\newcommand{\be}{\begin{equation}}
\newcommand{\ee}{\end{equation}}
\newcommand{\beq}{\begin{equation}}
\newcommand{\eeq}{\end{equation}}
\newcommand{\ben}{\begin{displaymath}}
\newcommand{\een}{\end{displaymath}}
\newcommand{\beqa}{\begin{eqnarray}}
\newcommand{\eeqa}{\end{eqnarray}}
\newcommand{\bea}{\begin{eqnarray}}
\newcommand{\eea}{\end{eqnarray}}
\newcommand{\bean}{\begin{eqnarray*}}
\newcommand{\eean}{\end{eqnarray*}}
\newcommand{\ba}{\begin{array}}
\newcommand{\ea}{\end{array}}
\newcommand{\bi}{\begin{itemize}}
\newcommand{\ei}{\end{itemize}}

\newcommand{\ie}{{\it i.e.,\,}}
\newcommand{\eg}{{\it e.g.,\,}}

\newcommand{\bbr}[1]{\mbox{${\mathbb R}^{#1}$}}

{\def\lesssim{\mathrel{\mathpalette\vereq<}}}{}
{}{}
\def\vereq#1#2{\lower3pt\vbox{\baselineskip1.5pt \lineskip1.5pt
\ialign{$\m@th#1\hfill##\hfil$\crcr#2\crcr\sim\crcr}}}
\makeatother

\begin{document}

\begin{flushright} CPHT-RR-009.0109 \end{flushright}\vskip .3in

\title{\font\titlerm=cmr10 scaled\magstep4\centerline{\titlerm
World-Volume Effective Theory}
\centerline{\titlerm
for Higher-Dimensional Black
Holes}
{(Original title: ``Blackfolds"\footnote{Modified to match the version
published in Phys.\ Rev.\ Lett.\  {\bf 102}, 191301 (2009).})}
\vspace{0.15cm}}
        
\author{Roberto Emparan$^{a,b}$, Troels Harmark$^{c}$,
Vasilis Niarchos$^{d}$,
Niels A.\ Obers$^{c}$ \\}
\address{\vspace{0.2cm} 
$^a$\,Instituci\'o Catalana de Recerca i Estudis
Avan\c cats (ICREA)\\ 
$^b$\,Departament de F{\'\i}sica Fonamental - Universitat de Barcelona,
Marti i Franqu\`es 1 E-08028 Barcelona, Spain \\
$^c$\,The Niels Bohr Institute, Blegdamsvej 17, 2100 Copenhagen \O, Denmark \\
$^d$\,Centre de Physique Th\'eorique - \'Ecole Polytechnique,
91128 Palaiseau, France, Unit\'e mixte de Recherche 7644, CNRS\\
{\small \tt emparan@ub.edu,
harmark@nbi.dk, niarchos@cpht.polytechnique.fr,
obers@nbi.dk} \\
}
\begin{abstract}
We argue that the main feature behind novel properties of
higher-dimensional black holes, compared to four-dimensional ones, is
that their horizons can have two characteristic lengths of very
different size. We develop a long-distance worldvolume effective theory
that captures the black hole dynamics at scales much larger than the
short scale. In this limit the black hole is regarded as a blackfold: a
{\it black} brane (possibly boosted locally) whose worldvolume spans a
curved submani{\it fold} of the spacetime. This approach reveals black
objects with novel horizon geometries and topologies more complex than
the black ring, but more generally it provides a new organizing
framework for the dynamics of higher-dimensional black holes.
\end{abstract}
\maketitle

\section{Introduction}

It has been realized in recent years that the dynamics of black holes in
spacetimes of dimension $D\geq 5$ is much richer than in four
dimensions. While the techniques developed to construct and characterize
four-dimensional black holes have been quite successful in five
dimensions, the dynamics in $D\geq 6$ seems to be too complex to be
captured by conventional approaches.

The main novel feature of higher-dimensional neutral black holes is that
in some regimes their horizons are characterized by at least two
separate scales, $r_0\ll R$. This does not occur in four dimensions,
where the shape of a Kerr black hole is always approximately round with
radius $r_0\sim GM$. In particular the angular momentum bound $J\leq
GM^2$ implies that rotation cannot produce large distortions on the
horizon. However, in $D\geq 5$ such bounds do not generally hold and the
two classical length scales $J/M$ and $(GM)^{1/(D-3)}$ can be widely
separated, as we know from explicit solutions. Five-dimensional black
rings can have arbitrarily large angular momentum for a given mass, and
at large $J$ the ring's radius $R$ is much bigger than its thickness
$r_0$ \cite{Emparan:2001wn}. Such black rings are also well-approximated
by (boosted) black strings. Likewise, in $D\geq 6$ Myers-Perry black
holes have ultra-spinning regimes with pancaked horizons approaching
black membranes of small thickness $r_0$ and large extent $R$
along the plane of rotation \cite{Myers:1986un,Emparan:2003sy}. There
are other phenomena peculiar to higher-dimensional horizons that depend
on the ability to separate two length scales along the horizon: the
Gregory-Laflamme instability and its associated inhomogeneous black
branes \cite{Gregory:1993vy} arise when the two scales characterizing
the thickness and the length of a black brane begin to differ. In
hindsight, it is surprising that four-dimensional black hole horizons
only possess short-scale ($\sim r_0$) dynamics. Thus it is clear that
new tools are needed in order to capture the long-distance ($\sim R\gg
r_0$) dynamics of higher-dimensional horizons.

The natural approach when faced with a problem with two widely separate
length scales is to integrate out the short-distance physics to obtain
a long-distance effective theory. In General Relativity there are two
different (but essentially equivalent) techniques to do this: matched
asymptotic expansions \cite{Harmark:2003yz} and the classical effective
field theory of \cite{Goldberger:2004jt}. To the order that we work in
this paper there is no difference between them. We shall refer to this
leading order theory for the long-distance dynamics of
higher-dimensional black holes as the blackfold approach. This program
was initiated in \cite{Emparan:2007wm} with the construction of thin
black rings in $D\geq 5$. Here we present a general framework. 

For spacetime dimension $D$ and spatial dimension $p$ of the blackfold
worldvolume, we denote $n=D-p-3$. Choosing units where $16\pi
G=\Omega_{n+1}$ simplifies some equations.

\section{Blackfold equations}

A blackfold is a {\em black} $p$-brane whose worldvolume extends along a
curved submani{\em fold} of the embedding spacetime. Beginning from a
flat black $p$-brane with horizon $\bbr{p}\times s^{n+1}$ (we denote the
sphere of short-size $r_0$ with lowercase $s$), we imagine bending
its worldvolume $\bbr{p,1}$ into some submanifold
$\mathcal{W}_{p+1}$, with spatial section $\mathcal{B}_p$
compact or not, characterized by a length-scale (\eg intrinsic
curvature radius) $R\gg r_0$. Describing the long-distance dynamics of
the blackfold amounts to finding the spacetime embedding
$X^\mu(\sigma^\alpha)$ of its worldvolume. This embedding determines the
induced (pulled-back) metric, $\gamma_{\alpha\beta}$,
and the first fundamental tensor $h_{\mu\nu}$, which acts as a projector
onto the worldvolume,
\beq
\gamma_{\alpha\beta}=g_{\mu\nu}\partial_\alpha X^\mu \partial_\beta
X^\nu\,,
\quad
h^{\mu\nu}=\gamma^{\alpha\beta}\partial_\alpha X^\mu \partial_\beta
X^\nu
\,.
\eeq

A general theory of the classical dynamics of a brane-like object,
regarded as a source of energy-momentum $T_{\mu\nu}$ in the probe
approximation, was developed in \cite{Carter:2000wv}. The consistent coupling
of the source to gravity requires $\nabla_\mu T^{\mu\nu}=0$. This
implies that 
\beq \label{carter}
T^{\mu\nu}K_{\mu\nu}{}^\rho=0 \,,
\eeq 
where
$K_{\mu\nu}{}^\rho$ is the extrinsic curvature tensor of the blackfold
embedding, 
\beq
K_{\mu\nu}{}^\rho=h^\lambda{}_\mu h^\sigma{}_\nu
\nabla_\lambda h^\rho{}_\sigma\,,
\eeq
with indices $\mu,\nu$ along directions tangent to the brane worldvolume
and
$\rho$ orthogonal to it. 

The effective stress-energy tensor of a blackfold is determined by
matching to short-distance physics. We demand that locally, \ie on
scales $r\ll R$, the blackfold be equivalent to a black $p$-brane up to
a position-dependent Lorentz-transformation. The gravitational field of
such a black $p$-brane is known, and at distances much larger than its
thickness, $r\gg r_0$, the field is weak. Thus we can determine an
equivalent distributional stress tensor
\beq
T_{\mu\nu}(\sigma^\alpha)=\tau_{\mu\nu}(\sigma^\alpha)\delta^{(D-p-
1)}\left(x-X(\sigma^\alpha)\right)
\eeq 
that
sources the same linearized field in the matching range $r_0\ll r\ll R$.

In \cite{toappear} we show that the blackfold equations (\ref{carter}) are
equivalent to a generalized geodesic equation
\begin{equation}
\label{newcarter}  
\tau^{\alpha \beta} \Big( \nabla^{(\gamma)}_\alpha
\partial_\beta X^\rho + \Gamma^\rho_{\mu \nu} \partial_\alpha X^\mu
\partial_\beta X^\nu \Big) = 0
\end{equation}
and also that they can be derived by coupling the worldvolume action
\beq \label{actionaa}
I [X^\mu(\sigma^\alpha)]=\int_{\mathcal{W}_{p+1}} dV_{p+1}~
\tau^{\alpha\beta}\gamma_{\alpha\beta}\, 
\eeq
to bulk gravity.

The effective stress tensor $\tau_{ab}$ of a {\it static} flat black
$p$-brane with orthonormal worldvolume coordinates $z^a=(z^0,z^i)$ is
\beq\label{tauab}
\tau_{00}=r_0^n (n+1)\,,\qquad \tau_{ii}=-r_0^n\,,
\eeq
where $r_0$ is the horizon radius in directions transverse to the
worldvolume, \ie the {\it thickness} of the black $p$-brane.

Consider now a Lorentz transformation $\Lambda \in SO(1,m)
\subset SO(1,p)$ of the $p$-brane, where $m \leq p $ is the number of
directions along which the blackfold is boosted, and which cannot be
larger than the number of independent rotation planes of
the spacetime. We parametrize the $m$ boosts as
\begin{equation}
\Lambda_0{}^0 = \cosh\alpha\,, \quad \Lambda_i{}^0 =
\nu_i \sinh \alpha\,, \quad
\sum_{i=1}^m \nu_i^2 = 1\,,
\end{equation}
where $\nu_i$ are the director cosines of a $S^{m-1}$ in parameter
space. The remaining components of $\Lambda$ are constrained by
$(\Lambda\eta\Lambda^T)_{ab}=\eta_{ab}$ but are
otherwise irrelevant.

We parametrize the background spacetime using coordinates
\beq
(t,r_1,\phi_1,\dots,r_m,\phi_m, x_1,\dots,x_{D-1-2m})\,,
\eeq 
where
$(r_l,\phi_l)$ are polar coordinates for the $l$-th rotation plane, with
$r_l$ measuring proper distance along the orbits of $\partial_{\phi_l}$.
The embedding is specified by 
\beq
X^\mu(\sigma^\alpha)=\left(t(\sigma^\alpha), r_l(\sigma^\alpha),
\phi_l(\sigma^\alpha), x_k(\sigma^\alpha)\right). 
\eeq
We specialize from now on to static
backgrounds and stationary blackfolds, \ie we
align $t \propto z^0=\sigma^0$ and take $X^i$ independent of $\sigma^0$.
We assume that $\partial_{\phi_l}$ generate isometries of the background
and align the $m$ boosted spatial coordinates on the $p$-brane with the
angular directions $\phi_l$. The $m$ boosted spatial coordinates are
thus identified periodically as $z^l\sim z^l+2\pi r_l(\sigma^\alpha)$.

In a blackfold the thickness and boost parameters in general depend on
the position, $r_0(\sigma^\alpha)$, $\alpha (\sigma^\alpha)$,
$\nu_i(\sigma^\alpha)$. However, to ensure regularity of the
horizon we impose a {\em blackness condition}: the surface gravity
$\kappa$ and the angular velocities $\Omega_{\mathrm{H}i}$ must be uniform over
$\mathcal{B}_p$. Locally, these are determined by the horizon properties
of a boosted black $p$-brane,
\beq
\kappa=\frac{n}{2 r_0(\sigma^\alpha) \cosh\alpha(\sigma^\alpha)}\,,\quad 
\Omega_{\mathrm{H}i}=
\frac{\nu_i(\sigma^\alpha)}{r_i(\sigma^\alpha)}\tanh\alpha(\sigma^\alpha)\,.
\eeq
Requiring blackness
determines the thickness and
boosts in terms of the local velocity components
$r_i(\sigma^\alpha)\Omega_{\mathrm{H}i}$,
\begin{equation}
\tanh \alpha (\sigma^\alpha) = \Xi (\sigma^\alpha)
\,,\quad
r_0 (\sigma^\alpha) = \frac{n}{2 \kappa} \sqrt{1- \Xi(\sigma^\alpha)^2}
\,,\quad
\nu_i (\sigma^\alpha) = \frac{r_i(\sigma^\alpha)\Omega_{\mathrm{H}i}}{\Xi(\sigma^\alpha)}\,, 
\end{equation}
where the local velocity field is
\begin{equation}
\label{Xis}
\Xi (\sigma^\alpha)=\left( \sum_{i=1}^m \left(r_i(\sigma^\alpha)\Omega_{\mathrm{H}i}\right)^2
\right)^{1/2}\,.
\end{equation}
Boosting $z^a\to (\Lambda z)^a$, $\tau_{a b}
\to (\Lambda \tau \Lambda^T)_{ab}$ we obtain
\begin{equation}
\label{tautt}
\tau_{00} =  
 \left( \frac{n}{2 \kappa} \right)^n ( 1- \Xi^2)^{\frac{n-2}{2}} 
\left(n+1 -\Xi^2\right)\,,
\end{equation}
\begin{equation}
\label{tauti}
\tau_{0i} =  
 \left( \frac{n}{2 \kappa} \right)^n  ( 1- \Xi^2)^{\frac{n-2}{2}} n  r_i
\Omega_{\mathrm{H}i}\,,
 \quad i = 1, \ldots, m
\end{equation}
\beq \label{tauii}
\tau_{ii}=\left( \frac{n}{2 \kappa} \right)^n  
( 1- \Xi^2)^{\frac{n}{2}} 
\left(\frac{n (r_i \Omega_{\mathrm{H}i})^2}{1 - \Xi^2}- 1\right)\,,
 \quad i = 1, \ldots, m
\eeq
\beq \label{tauii2}
\tau_{ii}=-\left(\frac{n}{2 \kappa} \right)^n  ( 1-
\Xi^2)^{\frac{n}{2}}\,,
 \quad i = m+1, \ldots, p
\eeq
\begin{equation}
\label{tauij}
\tau_{i\neq j } =   
 \left( \frac{n}{2 \kappa} \right)^n (1- \Xi^2)^{\frac{n-2}{2}}
r_i r_j \Omega_{\mathrm{H}i}\Omega_{\mathrm{H}j}\,,
   \quad i,j = 1, \ldots, m\,.
\end{equation}

As a consequence of the blackness condition, once the
$\Omega_{\mathrm{H}i}$ are given the blackfold equation \eqn{carter} becomes a purely
geometric one involving only $X^\mu(\sigma^\alpha)$ and its first and second
derivatives. $\kappa$
factorizes out of the equations and enters only to fix the horizon
thickness $r_0(\sigma^\alpha)$.

The mass and angular momenta of the blackfold are obtained by
integrating the energy and momentum densities over $\mathcal{B}_p$,
on which $\partial_t= N\partial_{z^0}$ ($N$ accounts for a possible redshift
between the blackfold and infinity) and 
$\partial_{\phi_i}=r_i\partial_{z^i}$. Then
\beq\label{MJ}
M=\int_{\mathcal{B}_p}dV_{p}\; N\tau_{00}\,,\qquad
J_i=\int_{\mathcal{B}_p}dV_{p}\;r_i
\tau_{0i}\,.
\eeq

At each point on $\mathcal{B}_p$ we assume the presence
of a small sphere $s^{n+1}$ ($n\geq 1$) locally 
equal to that in a boosted black $p$-brane and so with area
\beq
a_\mathrm{H}(\sigma^\alpha)=\Omega_{n+1}r_0^{n+1}(\sigma^\alpha)\cosh\alpha(\sigma^\alpha)
=\Omega_{n+1} \left( \frac{n}{2 \kappa} \right)^{n+1} ( 1- \Xi^2)^{\frac{n}{2}}\,.
\eeq

The horizon of the blackfold is therefore a fibration of $s^{n+1}$ over
$\mathcal{B}_p$. If the fiber is regular everywhere then the topology of
the horizon is (topology of $\mathcal{B}_p$)$\times S^{n+1}$. However,
the size $r_0(\sigma^\alpha)$ will decrease to zero at `walls' on
$\mathcal{B}_p$ where the local
velocity approaches lightspeed $\Xi(\sigma^\alpha)\to 1$. If as a result
$\mathcal{B}_p$ has the topology of a $p$-ball, with the $s^{n+1}$
shrinking to zero size at $\partial\mathcal{B}_p$, then the horizon
topology is $S^{n+p+1}=S^{D- 2}$. 

The total area of the horizon is
\beq\label{AH}
A_\mathrm{H}=\int_{\mathcal{B}_p}dV_{p}\;a_\mathrm{H}(\sigma^\alpha)\,.
\eeq
The first law of black hole mechanics 
\beq
\delta M=\frac{\kappa}{8\pi G}\delta A_\mathrm{H} +\Omega_{\mathrm{H}i} \delta J_i 
\eeq
can be seen to be satisfied by
solutions of the blackfold equations \cite{toappear}. 

\section{Explicit solutions}


\subsection{Products of odd-spheres}

Consider first a single
odd-sphere $S^{2k+1}$, which we embed in a $2k+2$-dimensional flat
subspace of $\bbr{D-1}$ with metric
\beq
d\rho^2+\rho^2\sum_{i=1}^{k+1}\left(d\mu_i^2+\mu_i^2d\phi_i^2\right)\,,\qquad
\sum_{i=1}^{k+1}\mu_i^2=1\,.
\eeq
The sphere is embedded as $\rho=R$ and the worldvolume spatial
coordinates can be taken to be $k$ independent $\mu_i$ plus the $k+1$
Cartan angles $\phi_i$. Then we have
$r_i=R\mu_i$. Assume now that all the angular velocities along the
$\phi_i$ are equal in magnitude, $|\Omega_{\mathrm{H}i}|=\Omega_\mathrm{H}$. From
(\ref{Xis}) the boost velocity $\Xi=R\Omega_\mathrm{H}$ is uniform over the
blackfold, and so is the thickness $r_0$. The blackfold equilibrium
equations easily reduce to
\beq
R=\sqrt{\frac{p}{n+p}}\frac{1}{\Omega_\mathrm{H}}
\eeq
(for $p=1$ we recover the result for black rings in \cite{Emparan:2007wm}). The
horizon geometry is $\mathcal{H}=S^{2k+1}\times s^{n+1}$. One may also
consider non-equal angular velocities. Then the radius $\rho=R(\mu_i)$
depends non-trivially on $\mu_i$ and one must solve a second-order
differential equation, which requires numerical analysis.

Consider now a product of odd-spheres,
${\mathcal{B}_p}=\prod_{p_a\in \mathrm{odd}} S^{p_a}$, $p=\sum_a p_a$,
embedded in a
flat subspace of $\bbr{D-1}$
with metric
\beq
\sum_a \left(d\rho_a^2+\rho_a^2 d\Omega_{p_a}^2\right)\,.
\eeq
Clearly the total number of spheres cannot be larger
than $D-1-p=n+2$. We look for blackfold geometries at constant radii
$\rho_a=R_a$, with each odd-sphere rotating along all its Cartan
angles with angular velocities equal in magnitude to $\Omega_{\mathrm{H}}^{(a)}$.
The equations of equilibrium factorize for each sphere and are solved
for
\beq
R_a=\sqrt{\frac{p_a}{n+p}}\frac{1}{\Omega_{\mathrm{H}}^{(a)}}\,.
\eeq
The horizon geometry is $
\mathcal{H}=\prod_{p_a\in \mathrm{odd}} S^{p_a}\times s^{n+1}$, and
the mass, angular momenta, and area of the blackfold are easily obtained
plugging these results in the general formulas above.

\subsection{Ultraspinning Myers-Perry black holes as even-ball blackfolds} 

The blackfold equations in a Minkowski background do not
admit solutions for ${\mathcal{B}_{p}}$ a topological even-sphere --- the
tension at fixed-points of the rotation group cannot be counterbalanced
by centrifugal forces. Instead they admit solutions where
${\mathcal{B}_{p}}$ is an ellipsoidal even-ball, with thickness $r_0$
vanishing at the boundary of the ball so the horizon topology is
$S^{D-2}$. These reproduce {\it precisely all} the physical properties
of a Myers-Perry black hole with $p/2$ ultra-spins, which provides a
highly non-trivial check on the approach. It also shows that the method
remains sensible when the rotation has fixed-points, in this case at the
center of the ball. They also exemplify blackfolds with varying
thickness $r_0(\sigma^\alpha)$.

We illustrate these solutions in the simplest non-trivial
case of $p=2$, and postpone the general analysis to \cite{toappear}.
Consider a black 2-fold extending along a plane 
\beq
dr^2+r^2 d\phi^2
\eeq
in Minkowski space. Being a plane, this ${\mathcal{B}_2}$ solves
trivially the blackfold equations \eqn{carter}. To set the blackfold in
rotation along the $\phi$ axis we embed
${\mathcal{B}_2}$ as $\sigma^1=\phi$, $\sigma^2=r$, with local boost
along $\sigma^1$, and obtain
$\Xi=r\Omega_\mathrm{H}$. Uniform $\Omega_\mathrm{H}$, \ie rigid
brane rotation, makes the local boost become lightlike at
$r=\Omega_\mathrm{H}^{-1}$. Constancy of $\kappa$ implies that $r_0(r\to
\Omega_\mathrm{H}^{-1})\to 0$ so ${\mathcal{B}_2}$ becomes the disk $0\leq r\leq
\Omega_\mathrm{H}^{-1}$. The physical magnitudes of the blackfold
are\footnote{These formulas correct the typos in the
journal version in Phys.\ Rev.\ Lett.\  {\bf 102}, 191301 (2009).}
\beqa
M&=&2\pi\left(\frac{n}{2\kappa}\right)^n\frac{1}{\Omega_\mathrm{H}^2}\frac{n+3}{n+2}\,,\qquad
J=2\pi\left(\frac{n}{2\kappa}\right)^n\frac{1}{\Omega_\mathrm{H}^3}\frac{2}{n+2}\,,\nonumber\\
A_\mathrm{H}&=&
2\pi\,\Omega_{n+1}\left(\frac{n}{2\kappa}\right)^{n+1}\frac{1}{\Omega_\mathrm{H}^2}
\frac{1}{n+2}\,.
\eeqa
If we now write $\Omega_\mathrm{H}=a^{-1}$ and $\frac{n}{2\kappa}=r_+$,
restore Newton's $G$ and use the identity $2\pi
\Omega_{n+1}=(n+2)\Omega_{n+3}$, these equations reproduce {\it exactly}
the values for an ultraspinning Myers-Perry black hole in $D=n+5$
dimensions, with a single spin parameter $a$ and with horizon radial
coordinate $r_+$, to leading order in $r_+/a$ \cite{Emparan:2003sy}. The
shape of the horizon is also accurately reproduced: for the
ultraspinning black hole the thickness transverse to the rotation plane
is $r_+\cos\theta$ \cite{Emparan:2003sy}, while for the blackfold,
introducing $\theta=\arcsin(\Omega_\mathrm{H} r)$, we find thickness
$r_0(\theta)=\frac{n}{2\kappa}\cos\theta$, thus in perfect agreement.
Also in both cases $a$ is the horizon radius in the plane parallel to
the rotation. Observe that once the angular velocity and surface gravity
are identified in the two constructions, there is no ambiguity when
comparing physical magnitudes.

\subsection{Minimal blackfolds}

For static blackfolds with space components of the stress tensor
$\tau^i{}_j=-P\delta^i{}_j$, the blackfold equations reduce to
\beq
K^\rho=0\,,
\eeq
where $K^\rho=h^{\mu\nu}K_{\mu\nu}{}^\rho$ is the mean curvature vector.
Thus, ${\mathcal{B}_p}$ must span a (sufficiently regular and
non-self-intersecting) minimal spatial submanifold. As far as we know,
in Euclidean space no compact examples of these have been found. 

\section{Discussion and outlook}

Let us address some caveats about the blackfold approach. (I) One may
worry whether the horizon of the black brane remains regular after
bending its worldvolume. For black 1-folds, \ie thin black rings,
ref.~\cite{Emparan:2007wm} showed that this is the case iff the
blackfold equations \eqn{carter} are satisfied. An extension to black
$p$-folds will be given elsewhere \cite{toappear}. (II) To leading order
in $r_0/R$ the backreaction of the blackfold on the background geometry
is neglected. It may happen that backreaction makes it impossible for a
leading-order solution to remain stationary --- by developing naked
singularities revealing unbalanced forces, or more physically, by
inducing evolution in time. This must be analyzed in a case-by-case
basis, typically using physical input about the expected effects of
self-gravitational attraction. Many minimal blackfolds
likely exhibit this phenomenon. Such solutions would not correspond to
actual stationary black holes, but they would still be of interest as
blackfolds that evolve slowly (at least initially) with timescale small
in $r_0/R$. (III) Blackfolds may be (classically) unstable. Stability to
long-wavelength $(\lambda\gg r_0)$ perturbations can be analyzed using
the blackfold equations. There are however short-wavelength
$(\lambda\sim r_0)$ instabilities, \eg of Gregory-Laflamme type,
outside the approach which would proceed on quick timescales,
$\Gamma\sim 1/r_0$. 

We have presented the theory of neutral blackfolds, and examples have
referred to stationary blackfolds in a Minkowski background. However,
the method can be readily generalized \cite{toappear} to charged
blackfolds as well as other backgrounds (\eg \cite{Caldarelli:2008pz})
and some blackfold motions. 

Our approach naturally organizes the dynamics of higher-dimensional
neutral black holes according to the relative value of the lengths
$(GM)^\frac{1}{D-3}$ and $J/M$, where $J=\left(\sum_i
J_i^2\right)^{1/2}$:

(i) $0\leq J\lesssim M(GM)^\frac{1}{D-3}$: there is a single length
scale on the horizon and the physics is qualitatively similar to the
Kerr black hole.

(ii) $J \sim M(GM)^\frac{1}{D-3}$: regime of mergers and connections
between phases that occur when the two horizon scales meet, $r_0\sim R$.
Such threshold phenomena occur outside the limit of validity of
effective field theories and are very difficult to analyze, requiring
extrapolations, new approaches or numerical analysis (also, some mergers
may occur at large values of $J$ \cite{Emparan:2007wm}).

(iii) $J\gg M(GM)^\frac{1}{D-3}$: blackfolds. We have developed the tools
to study the extremely rich physics in this regime. Rather than search
for exact solutions for {\it all} possible higher-dimensional black
holes, it seems to us more fruitful to study the dynamical features of
blackfolds.

\begin{acknowledgments}
We thank Elias Kiritsis and Barak Kol for useful discussions. Warm
hospitality at different stages in this project was provided by TIFR and
ICTS during the 2008 Monsoon Workshop on String Theory in Mumbai, and
the Cosmophysics group in KEK, Tsukuba (RE), the Galileo Galilei
Institute in Firenze (TH, NO), Niels Bohr Institute (RE, VN), \'Ecole
Polytechnique (NO), and the CERN TH Institute program on Black Holes
(all authors). RE was supported by DURSI 2005 SGR 00082, MEC FPA
2007-66665-C02 and CPAN CSD2007-00042 Consolider-Ingenio 2010. TH was
supported by the Carlsberg foundation. VN was supported by an Individual
Marie Curie Intra-European Fellowship and by ANR-05-BLAN-0079-02 and
MRTN-CT-2004-503369, and CNRS PICS {\#} 3059, 3747 and 4172. All authors
were supported by MRTN-CT-2004-005104.
\end{acknowledgments}


\begin{thebibliography}{99}

\bibitem{Emparan:2001wn}
  R.~Emparan and H.~S.~Reall,
  ``A rotating black ring in five dimensions,''
  Phys.\ Rev.\ Lett.\  {\bf 88} (2002) 101101
  [arXiv:hep-th/0110260].

  R.~Emparan and H.~S.~Reall,
  ``Black rings,''
  Class.\ Quant.\ Grav.\  {\bf 23} (2006) R169
  [arXiv:hep-th/0608012].

\bibitem{Myers:1986un}
  R.~C.~Myers and M.~J.~Perry,
  ``Black Holes In Higher Dimensional Space-Times,''
  Annals Phys.\  {\bf 172} (1986) 304.


\bibitem{Emparan:2003sy}
  R.~Emparan and R.~C.~Myers,
  ``Instability of ultra-spinning black holes,''
  JHEP {\bf 0309} (2003) 025
  [arXiv:hep-th/0308056].


\bibitem{Gregory:1993vy}
  R.~Gregory and R.~Laflamme,
  ``Black strings and p-branes are unstable,''
  Phys.\ Rev.\ Lett.\  {\bf 70}, 2837 (1993)
  [arXiv:hep-th/9301052].

  T.~Harmark, V.~Niarchos and N.~A.~Obers,
  ``Instabilities of black strings and branes,''
  Class.\ Quant.\ Grav.\  {\bf 24} (2007) R1
  [arXiv:hep-th/0701022].


\bibitem{Harmark:2003yz}
  T.~Harmark,
  ``Small black holes on cylinders,''
  Phys.\ Rev.\  D {\bf 69} (2004) 104015
  [arXiv:hep-th/0310259].

  D.~Gorbonos and B.~Kol,
  ``A dialogue of multipoles: Matched asymptotic expansion for caged black holes,''
  JHEP {\bf 0406}, 053 (2004)
  [arXiv:hep-th/0406002].

\bibitem{Goldberger:2004jt}
  W.~D.~Goldberger and I.~Z.~Rothstein,
  ``An effective field theory of gravity for extended objects,''
  Phys.\ Rev.\  D {\bf 73}, 104029 (2006)
  [arXiv:hep-th/0409156].

  Y.~Z.~Chu, W.~D.~Goldberger and I.~Z.~Rothstein,
  ``Asymptotics of d-dimensional Kaluza-Klein black holes: Beyond the newtonian approximation,''
  JHEP {\bf 0603}, 013 (2006)
  [arXiv:hep-th/0602016].

  B.~Kol and M.~Smolkin,
  ``Classical Effective Field Theory and Caged Black Holes,''
  Phys.\ Rev.\  D {\bf 77}, 064033 (2008)
  [arXiv:0712.2822 [hep-th]].

\bibitem{Emparan:2007wm}
  R.~Emparan, T.~Harmark, V.~Niarchos, N.~A.~Obers and M.~J.~Rodriguez,
  ``The Phase Structure of Higher-Dimensional Black Rings and Black Holes,''
  JHEP {\bf 0710}, 110 (2007)
  [arXiv:0708.2181 [hep-th]].



\bibitem{Carter:2000wv}
  B.~Carter,
  ``Essentials of classical brane dynamics,''
  Int.\ J.\ Theor.\ Phys.\  {\bf 40}, 2099 (2001)
  [arXiv:gr-qc/0012036].


\bibitem{toappear}
R.~Emparan, T.~Harmark, V.~Niarchos, N.~A.~Obers, to appear.

\bibitem{Caldarelli:2008pz}
  M.~M.~Caldarelli, R.~Emparan and M.~J.~Rodriguez,
  ``Black Rings in (Anti)-deSitter space,''
  JHEP {\bf 0811} (2008) 011
  [arXiv:0806.1954 [hep-th]].

  J.~Camps, R.~Emparan, P.~Figueras, S.~Giusto and A.~Saxena,
  ``Black Rings in Taub-NUT and D0-D6 interactions,''
  JHEP {\bf 0902} (2009) 021
  [arXiv:0811.2088 [hep-th]].




\end{thebibliography}
\end{document}